\documentclass[conference,final,twocolumn]{IEEEtran}
\usepackage{amsmath,amssymb,amsfonts,mathrsfs,bm}
\usepackage{amstext}
\usepackage{upgreek}
\usepackage{multicol}
\usepackage{graphicx}
\usepackage{paralist}
\usepackage{hyperref}
\usepackage[numbers,sort&compress]{natbib}
\usepackage{booktabs}
\usepackage{multirow}
\usepackage{subfigure}

\usepackage{algorithm}
\usepackage{algorithmic}
\newtheorem{proposition}{Proposition}

\IEEEoverridecommandlockouts

\begin{document}
\title{Caching Policy Optimization for D2D Communications by Learning User Preference}
\author{

\IEEEauthorblockN{{Binqiang Chen and Chenyang Yang }}
\IEEEauthorblockA{Beihang University, Beijing, China\\
Email: \{chenbq,cyyang\}@buaa.edu.cn}

\thanks{This work is supported by NSF Grants with No. 61671036  and 61429101.}
%
}
\maketitle

\begin{abstract}
Cache-enabled device-to-device (D2D) communications can boost network throughput. By pre-downloading contents to local caches of users, the content requested by a user can be transmitted via D2D links by other users in proximity. Prior works optimize the caching policy at users with the knowledge of \emph{content popularity}, defined as the probability distribution of request for every
file in a library from by all users. However, content popularity can not reflect the interest of each individual user and thus popularity-based caching policy may not fully capture the performance gain introduced by caching.  In this paper, we optimize caching policy for cache-enabled D2D by learning \emph{user preference}, defined as the conditional probability distribution of a user's request for a file given that the user sends a request. We first formulate an optimization problem with given user preference to maximize the offloading probability, which is proved as NP-hard, and then provide a greedy algorithm to find the solution. In order to predict the preference of each individual user, we model the user request behavior by probabilistic latent semantic analysis (pLSA), and then apply expectation maximization (EM) algorithm to estimate the model parameters. Simulation results show that the user preference can be learnt quickly. Compared to the popularity-based caching policy, the offloading gain achieved by the proposed policy can be remarkably improved even with predicted user preference.
\end{abstract}


\section{Introduction}

Device-to-device (D2D) communications can boost the throughput of cellular networks and is a promising way to support the exponential growth of traffic load of 5th generation (5G) wireless networks \cite{doppler2009device}.

Motivated by the fact that a few popular contents account for most of the traffic load, caching contents at the wireless edge has become a trend for content delivery \cite{Procach14,wang2014cache}. By proactively downloading contents to base stations (BSs) or users, both the network throughput and energy efficiency as well as the user experience can be improved dramatically \cite{Golrezaei.TWC,dongcaching16,Chen15}.

Considering the limited storage size at wireless edge compared to the huge number of available contents, proactive caching policy is critical to achieve the performance gain of local caching \cite{Procach14}. Inspired by such observation, all prior works optimize caching policies with the knowledge of {\em content popularity}, which can be defined as the probability of request for each content in a catalogue from all users in a certain region during a period of time. It can reflect the collective behavior of users in a community \cite{tatar2014survey}.

By assuming identical content popularity among different cells, a caching policy was optimized for each BS to minimize the average download delay with known BS-user topology in \cite{niki2012femto}, and a probabilistic caching policy for BS was proposed  in \cite{B2015optimal} considering the uncertainty of BS-user topology. Under the same assumption, a trade-off between outage and throughput of cache-enabled D2D communications was investigated and a probabilistic caching policy was optimized in \cite{JMY.JSAC}. Considering that different social groups may prefer different types of contents, selfish and cooperative caching policies for user groups with different content popularities were optimized  in \cite{guo2015cooperative} to minimize the average delay.

{ Content popularity} is assumed perfectly known in these works  \cite{niki2012femto,B2015optimal,JMY.JSAC,guo2015cooperative}, which however is unknown  in practice and needs to be predicted \cite{Gha2016provab}, either directly \cite{tatar2014survey} or indirectly via various machine learning algorithms. Prediction for content popularity in communities with large number of users, e.g., who visit Youtube website, has become an active research field recently. Many prediction methods have been proposed, such as using cumulative views statistics
based on the popularity correlation over time \cite{tatar2014survey} and learning with a multi-armed bandit algorithm in \cite{Blasco14learning}. Inspired by collaborative filtering \cite{ekstrand2011collaborative} and assumed that $80\%$ of the ratings for each file have been provided by each user, the $20\%$  unknown ratings was predicted by singular value decomposition in \cite{Procach14}. The predicted ratings for each file are then aggregated to reflect file popularity, and the files with higher total ratings were cached.

In fact, as a demand statistic of large amount of users, content popularity is not able to reflect the preference of each individual user, which can be defined as the probability that a particular content is requested by a specific user. Existing works in the literature of local caching do not differentiate user preference and content popularity, which implicitly assume that user preferences are identical among users in a region or in a social group, and equals to the content popularity. Such an assumption is invalid, noting that the user preferences are heterogeneous in reality, especially for a small region such as a cellular cell with small user population.

D2D communications is only applicable for users in proximity, and thus the number of users that each user can share files with is rather limited. On the other hand, collaborative filtering is a power tool to directly predict user preference \cite{ekstrand2011collaborative}, which has been extensively investigated in the applications for recommendation problem. This motivates us to investigate the caching policy for D2D by leveraging the user preference learned by collaborative filtering algorithms.

In this paper, we study optimal caching policy for cache-enabled D2D communications with unknown and heterogeneous user preferences. D2D communications is only allowed between users within a collaboration distance to ensure the high data rate at receiver and low energy cost at transmitter. We first formulate an optimization problem with given user preferences to maximize the probability that a user can find the requested file in its own cache or caches of other users in proximity, which is proved as NP-hard. To solve the problem efficiently, we develop a greedy algorithm. In order to learn and  exploit user preferences, we model user request behavior resorting to probabilistic latent semantic analysis (pLSA), which was originally developed for automatic indexing and information retrieval \cite{hofmann1999prob} and then became a probabilistic model in collaborative filtering \cite{ekstrand2011collaborative}. The model parameters are estimated by maximizing likelihood function using expectation maximization (EM) algorithm \cite{demp1977maximum}. Simulation results show that user preferences can be quickly learnt. Compared to simply regarding content popularity as user preference (either with perfect or predicted content popularity) as in priori works, the offloading gain can be remarkably increased by the proposed caching policy with the predicted user preference.

\section{System Model}
\label{sec:system model}
We consider a cell with  $K$ uniformly distributed users, which constitutes a user set $\mathcal{U} = \{{\rm u}_1,{\rm u}_2,...,{\rm u}_K\}$. The BS is connected to the core network via backhaul. Each single-antenna user has a local cache to store $M$ files, and can act as a helper to share files via D2D link.

To provide high rate  transmission with low energy cost, we consider a user-centric communication protocol. A helper can serve as a D2D transmitter and send its cached files to a user only if their distance is smaller than a given value $r_{\rm c}$, called {\em collaboration distance}. The BS is aware of the cached files at each user and coordinates the D2D communications.

\subsection{Content Popularity and User Preference}
Consider a static content library $\mathcal{F} = \{{\rm f}_1,{\rm f}_2,...,{\rm f}_{F}\}$ consisting of $F$ files that all users in the cell may request. Each file is with the same file size. The request statistics for all users and for each individual user are defined as follows.

{\em Content popularity} is the probability distribution of requests for the files in the library from all users in the cell, denoted as  ${\bf p} = [p_1,p_2,...,p_{F }]$, where $p_{f} = P({\rm f}_f)$ is the probability that the $f$th file is requested, $\sum_{f=1}^{F} p_f = 1$, $p_{f}\in [0,1]$, and $1 \leq f \leq F$. Content popularity reflects the common interests of all users in the cell.

{\em User preference} is the conditional probability distribution of a user's request for every file given that the user sends a request, denoted as ${\bf q}_k= [q_{k,1},q_{k,2},...,q_{k,F }]$ for the $k$th user, where $q_{k,f} = P({\rm f}_f|{\rm u}_k)$ is the conditional probability that the $k$th user will request the $f$th file if the user sends a file request, $\sum_{f=1}^{F} q_{k,f}= 1$, $q_{k,f}\in [0,1]$, $1 \leq f \leq F$ and $1 \leq k \leq K$. We use matrix $\textbf{Q}:= (q_{k,f})^{K \times F}$ to denote the preferences of all users, where $(q_{k,f})^{K \times F}$ represents a matrix with $K$ rows and $F$ columns and $q_{k,f}$ as the element in the $k$th row and $f$th column. User preference reflects the personal interest of each individual user.

Assume that both content popularity and user preference are stationary, then their relation is given by
\begin{equation}
	\label{p_f}\textstyle
	p_f = \sum_{k=1}^{K} w_k q_{k,f},
\end{equation}
where $w_k = P({\rm u}_k)$, which is the probability that a request is sent by ${\rm u}_k \in \mathcal{U}$, $w_k \in [0,1]$, and $\sum_{k=1}^{K} w_k = 1$. Denote ${\bf w} = [w_1,w_2,...,w_K]$, which is the request distribution and reflects the active level of each user.


\emph{Content popularity} is often modeled by a Zipf distribution. The probability that the $f$th file in the library is requested is
\begin{equation}
\label{content_p}\textstyle
p_f = {f^{-\beta}}/{\sum_{j=1}^{F} j^{-\beta} },
\end{equation}
where the files in the library are indexed in descending order of popularity, and the content popularity distribution is more skewed with larger $\beta$  \cite{{niki2012femto}}.

\emph{User preferences} model, unfortunately, is not available in the literature so far. To investigate the impact of optimizing caching policy by learning user preference, we synthesize the user preference from the content popularity, inspired by a method in \cite{leconte2016placing} to synthesize content popularity in a cell from that of a core network, with the steps as follows:
\begin{itemize}
	\item ${\rm u}_k$ is associated with a feature value $X_k$, which is a uniform random variable randomly selected from $[0,1]$.
	\item ${\rm f}_f$ is associated with a feature value $Y_f$, which is again chosen uniformly and independently from $[0,1]$.
	\item Define a kernel function to reflect the correlation between the $k$th user and the $f$th file as $g(X_k,Y_f) = (1-|X_k-Y_f|)^{(\frac{1}{\alpha^3}-1)}$, where $0<\alpha \leq 1$. Here $g(X_k,Y_f) $ takes values in $[0,1]$, where $g(X_k,Y_f) = 0$ means that the $f$th file will not be requested by the $k$th user, and $g(X_k,Y_f) = 1$ can be interpreted as that the $f$th file belongs to the preferred file type of the $k$th user.
\end{itemize}
Then, the joint probability that the $f$th file is requested by the $k$th user is given as
\begin{equation}
\label{user_p}\textstyle
P({\rm u}_k, {\rm f}_f) = w_kq_{k,f} = p_f \frac{g(X_k,Y_f)}{\sum_{k'=1}^{K} g(X_{k'},Y_f)}.
\end{equation}

The main differences with \cite{leconte2016placing} lie in that: (1) we use \eqref{user_p} to construct the joint probability instead of the number of requests for the $f$th file at a specific BS, and (2) we introduce a parameter $\alpha$ into the kernel function  to capture the similarity of preferences among all users rather than simply using $g(X_k,Y_f) = (1-2|X_k-Y_f|)^4$ in \cite{leconte2016placing}.

To understand the role of parameter $\alpha$ in the kernel function, we employ the notion of cosine similarity from collaborative filtering \cite{ekstrand2011collaborative}, which is frequently used to reflect the similarity level of preferences between two users and is defined as
\begin{equation}
\label{cos_sim}\textstyle
\text{sim}({\bf q}_k,{\bf q}_m) = \frac{ \sum_{f=1}^{F} q_{m,f} q_{k,f}}{\sqrt{\sum_{f=1}^{F} q_{m,f}^2 \sum_{f=1}^{F} q_{k,f}^2}}.
\end{equation}
To show the similarity among the preferences of all users, we can define average cosine similarity as
\begin{equation}
\label{ave_sim}\textstyle
\mathbb{E}_{k,m} [ \text{sim}({\bf q}_k,{\bf q}_m) ]= \frac{2}{K(K-1)}\sum_{k,m}\frac{ \sum_{f=1}^{F} q_{k,f} q_{m,f}}{ \sqrt{\sum_{f=1}^{F} q_{k,f}^2 \sum_{f=1}^{F} q_{m,f}^2}}.
\end{equation}

{\bf Remark 1:}
When $\alpha = 1$, $g(X_k,Y_f) = 1$ for $\forall k,f$. Then, all user preferences are identical and equal to the file popularity, and $\mathbb{E}_{k,m}\{\text{sim} ({\bf q}_k,{\bf q}_m) \}=1$. When $\alpha \rightarrow 0$, $g(X_k,Y_f) \rightarrow 0$ for $X_k \neq Y_f$, and $g(X_k,Y_f) = 1$ only for $X_k = Y_f$. This indicates that $g(X_k,Y_f)g(X_{k'},Y_f) \rightarrow 0, k \neq k'$. Then, no user has the same preference, and $\mathbb{E}_{k,m}[ \text{sim} ({\bf q}_k,{\bf q}_m) ] \rightarrow 0$. We can see that $\alpha$ in the kernel function can reflect the average cosine similarity among users.

\subsection{Content Delivery and Caching Placement}

In practice, user preferences are unknown and need to be learned based on accumulated user requests. Considering that the traffic load at a BS varies periodically \cite{paul2011understanding}, we can divide time into periods as in \cite{Blasco14learning}. Each period includes a peak time and an off-peak time. During each period, there are a {\em content delivery} phase and a {\em caching placement} phase.

In content delivery phase, each user requests files according to its own preference. If a user can find its requested file in local cache, it directly retrieves the file with zero delay. If not, but the user can find the file in caches of other users with distance smaller than $r_{\rm c}$, the user establishes a D2D link with the closet user cached the file to fetch it. Otherwise, the user accesses to the BS to fetch the file via backhaul. Both fetching locally and via D2D links are called fetching via D2D links in this paper, because fetching locally can be regraded as fetching via D2D with extremely high data rate. Let $\textbf{A}:= (a_{i,j})^{K \times K} \in \{0,1\}^{K\times K}$ represent the topology relation between users, where $a_{i,j}=1$ if the distance between the $i$th and $j$th users is smaller than $r_{\rm c}$ and $a_{i,j}=0$ otherwise. Denote file requests matrix after a period as ${\bf N}:=(n({\rm u}_k, {\rm f}_f))^{K \times F}$, where $n({\rm u}_k, {\rm f}_f) \ge 1$ is the cumulative number of requests from the user ${\rm u}_k \in \mathcal{U}$ for the ${\rm f}_f \in \mathcal{F}$ after the  period. 

In caching placement phase, the BS first predicts the user preferences $\bf {Q}$ and the request distribution ${\bf {w}}$ based on the requests history $\bf N$, and then optimizes the caching policy. We consider a deterministic caching policy, where the files placed at the $k$th user is denoted by a binary vector ${\bf c}_k = [c_{k,1},c_{k,2},...,c_{k,F}]$, $c_{k,f}=1$ if the $f$th file is cached at the $k$th user, $c_{k,f}=0$ otherwise, and $\sum_{f=1}^{F} c_{k,f}\leq M$. We assume that the topology between users remains fixed during the caching placement and delivery phases. When this assumption does not hold, probabilistic caching can be applied, while the optimized policy is not provided due to the space limitation.  Denote caching policy as $\textbf{C}:= (c_{k,f})^{K \times F}$, and the file set cached at the $k$th user as $\mathcal{C}_k = \{{\rm f}_f|c_{k,f}=1\}$. After optimizing ${\bf C}$, the BS refreshes the caches of users during the off-peak time of each period, where the energy cost for placing the files can be minimized by using the method in \cite{yao16energy}.

\section{Caching Policy Optimization}

In this section, we optimize the caching policy with given user preferences. To this end, we first obtain the offloading probability to reflect the offloading gain. Then, we formulate the optimization problem to maximize the offloading gain and show that it is NP-hard. Finally, we propose a greedy algorithm to solve the problem.

\subsection{Offloading Probability}

Once the requested file can be fetched via D2D links, the traffic at the BS can be offloaded and the user can enjoy high data rate and low delay. We use {\em offloading probability} to reflect the offloading gain introduced by cache-enabled D2D communications, which is defined as the probability that a user can fetch the requested file via D2D links.

Denote $\mathcal{U}_k \in \mathcal{U}$ as the user set that the $k$th user can establish D2D links with, where ${\rm u}_m \in \mathcal{U}_k$ if $a_{k,m} = 1$ and $\mathcal{U}_k = \{{\rm u}_m|a_{k,m}=1\}$. Denoting $\mathcal{F}_k\in \mathcal{F}$ as the file set that the $k$th user can fetch via D2D links, then $\mathcal{F}_k$ is the union of the cached contents at the users in $\mathcal{U}_k$, i.e.,
\begin{equation}
\begin{aligned}
\mathcal{F}_k & \textstyle= \cup_{{\rm u}_m \in \mathcal{U}_k } \mathcal{C}_m = \{{\rm f}_f | \sum_{{\rm u}_m \in  \mathcal{U}_k } c_{m,f} > 0\} \\
& \textstyle= \{{\rm f}_f | \sum_{m=1}^{K} a_{k,m}c_{m,f} > 0\}.
\end{aligned}
\end{equation}
Then, the offloading probability can be obtained as
\begin{equation}
\label{p_off}
\begin{aligned}
p_{\rm off} ({\bf Q,w,A,C}) &\textstyle= \sum_{k=1}^{K} w_k \sum_{{\rm f}_f \in \mathcal{F}_k} q_{k,f}  \\
&\textstyle=\sum_{k=1}^{K} w_k \sum_{f=1}^{F} q_{k,f} \text{sgn}\left(\sum_{m=1}^{K} a_{k,m}c_{m,f}\right) \\
&\textstyle\stackrel{(a)}{=} \sum_{f=1}^{F} \sum_{k=1}^{K} w_k q_{k,f} g_{k,f}({\bf A,C}),
\end{aligned}
\end{equation}
where $\text{sgn}(x) = \min\{\max\{0,x\},1\}$ denotes that $x$ is truncated by $0$ and $1$, and (a) comes from $g_{k,f}({\bf A,C}) \triangleq \text{sgn}\left(\sum_{m=1}^{K} a_{k,m}c_{m,f}\right) \leq 1$, which indicates whether the $f$th file can be fetched via D2D links by the $k$th user.

{\bf Remark 2:} Prior works assume known content popularity ${\bf p}$, and implicitly assume that all users send their requests with equal probability, and then the offloading probability is $p^{\rm pop}_{\rm off} ({\bf p,A,C})=\frac{1}{K} \sum_{f=1}^{F} p_{f} \sum_{k=1}^{K} g_{k,f}({\bf A,C})$.

{\bf Remark 3:} When the collaboration distance $r_c \rightarrow \infty$, $a_{k,m}=1$. Then, \eqref{p_off} can be rewritten as
\begin{equation}
\begin{aligned}\textstyle
p_{\rm off} ({\bf Q,w,A,C}) & \textstyle = \sum_{f=1}^{F} \text{sgn}\left(\sum_{m=1}^{K}c_{m,f}\right) \sum_{k=1}^{K} w_k q_{k,f} \\
& \textstyle = \sum_{f=1}^{F} \text{sgn}\left(\sum_{m=1}^{K}c_{m,f}\right)  p_f.
\end{aligned}
\end{equation}
In this extreme case, $p_{\rm off} ({\bf Q,w,A,C})=p^{\rm pop}_{\rm off} ({\bf p,A,C})$.



\subsection{Optimization for Caching Policy}
To maximize the offloading gain introduced by cache-enabled D2D communications, we optimize the caching policy by solving the following problem,
\begin{equation}
	\label{equ.opt1}
	\begin{aligned}\textstyle
	\textbf{P1}: \quad&  \textstyle \max_{c_{m,f}}\,\, && \textstyle p_{\rm off} ({\bf {Q},{w},A,C})  \\
	& \textstyle  s.t.  && \textstyle \sum_{f=1}^{F} c_{m,f} \leq M, c_{m,f} \in \{0,1\}, \\
	&&&  \textstyle 1 \leq m \leq K, 1 \leq f\leq F .
	\end{aligned}
\end{equation}

\begin{proposition}
	\label{proposition1}
	Solving \textbf{P1} is NP-hard.
\end{proposition}
\begin{IEEEproof}
The special case of problem \textbf{P1} when ${q}_{k,f} = {p}_f$ for $1 \leq k \leq K $ can be obtained as

\begin{equation}
\label{equ.opt2}
\begin{aligned}
\textbf{P2}: \quad&\textstyle \max_{c_{m,f}}\,\, && \textstyle\sum_{k=1}^{K} {w}_k \sum_{{\rm f}_f \in \mathcal{F}_k} {p}_{f}   \\
&  \textstyle s.t.  && \textstyle |\mathcal{C}_m| \leq M,  \mathcal{F}_k = \cup_{{\rm u}_m \in \mathcal{U}_k } \mathcal{C}_m, \\
&&& \textstyle1 \leq m \leq K, 1 \leq f\leq F,
\end{aligned}
\end{equation}
where $|\mathcal{C}_m|$ is the cardinality of $\mathcal{C}_m$. It is easy to show that \textbf{P2} has the same structure with the problem formulated in \cite{niki2012femto}, which has been proved as NP-hard \cite{niki2012femto}. Because \textbf{P2} is a special case of \textbf{P1}, problem \textbf{P1} is NP-hard.
\end{IEEEproof}

\subsection{Algorithm to Find the Caching Policy}
Since problem \textbf{P1} is NP-hard, it is impossible to find the optimal solution within polynomial time. We propose a greedy algorithm to solve the problem. It starts with zero elements for the caching matrix, i.e., $\textbf{C}:= (0)^{K \times F}$. In each step, the value of one element in \textbf{C} is changed from zero to one with the highest incremental caching gain defined as
\begin{equation}
\label{margin_f}
\begin{aligned}
& \textstyle v_{\bf C}(m,f) = p_{\rm off} ({{\bf {Q},{w},A,C}|_{c_{m,f}=1}}) - p_{\rm off} ({\bf {Q},{w},A,C})\\
& \textstyle\stackrel{(a)}{=}\sum_{k=1}^{K} w_k q_{k,f} \left( g_{k,f} \left({\bf A,}{\bf C}|_{c_{m,f}=1}\right)-g_{k,f}\left(\bf A,C\right)  \right),
\end{aligned}
\end{equation}
where $(a)$ follows by substituting \eqref{p_off}, $\bf C$ is the caching matrix at previous step, and ${\bf C}|_{c_{m,f}=1}$ is the matrix by letting $c_{m,f}=1$ in $\bf C$. The algorithm is summarized in Algorithm \ref{greedy_algo}.

\begin{algorithm}[!htb]
	\caption{Finding the solution of  problem \textbf{P1}.}
	\label{greedy_algo}
	\begin{algorithmic}[1] 
		\REQUIRE ~  
		 $\bf A$;
		 $\bf {w}$;
		 $\bf {Q}$; \\
		Initialize:
		Caching matrix $\textbf{C}:= (0)^{K \times F}$;
		Files not cached at the $m$th user $\mathcal{\bar C}_m \leftarrow \{{\rm f}_1,{\rm f}_2,...,{\rm f}_F\}$;
		Users with residual storage space $\mathcal{U}_0 \leftarrow \{{\rm u}_1,{\rm u}_2,...,{\rm u}_K\}$;
		
		\FOR{$i=1,2,...,K \times M$}
		\label{for_1}
		\STATE $[m^*,f^*] =\arg\max_{{\rm u}_m\in \mathcal{U}_0, {\rm f}_f\in\mathcal{\bar C}_m  } v_{\bf C}(m,f)$;  \\
		\label{margin_v}
		\STATE ${\bf C} := {\bf C}|_{c_{m^*,f^*}= 1} $;	 		
		\STATE $\mathcal{\bar C}_{m^*} \leftarrow \mathcal{\bar C}_{m^*}\setminus {\rm f}_{f^*} $;
		\IF{$|\mathcal{\bar C}_{m^*}| = F - M$}
		\STATE $\mathcal{U}_0 \leftarrow \mathcal{U}_0 \setminus {\rm u}_{m^*} $;
		\ENDIF
		\ENDFOR
		\STATE $\textbf{C}^* = {\bf C}$;
		\ENSURE Caching matrix $\textbf{C}^*$.
	\end{algorithmic}
\end{algorithm}

The loops in step \ref{for_1} of Algorithm \ref{greedy_algo} take  $KM$ iterations, because there are totally $KM$ files that can be cached at all users. The step \ref{margin_v} for finding the file with the highest incremental caching gain takes at most $KF$ iterations. For each time of finding the file, the time complexity is $O(K)$ according to \eqref{margin_f}. Hence the total time complexity for Algorithm \ref{greedy_algo} is $O(K^3MF)$.

{\bf Remark 4:} It is noteworthy that the solution of problem \textbf{P2} is a caching policy optimized with known content popularity. Algorithm \ref{greedy_algo} is also applicable for \textbf{P2} by letting  ${q}_{k,f} = {p}_f, \forall k,f$ in $\bf {Q}$.
Solutions based on Algorithm \ref{greedy_algo} for \textbf{P1} and \textbf{P2} are respectively called ${\textbf{S1}}$ and $\textbf{S2}$ in the sequel.

\section{Predicting User Preferences}
\label{sec:predict}
In this section, we strive to predict user preferences with a model-based collaborative filtering method by using the historical file requests of all users. We first use pLSA to model file request behavior of users. Then, we estimate parameters of pLSA with EM algorithm  by maximizing the likelihood.

\subsection{Modeling File Requests Behavior of Users}
PLSA was first developed as an approach for automatic indexing and information retrieve, which provides a probabilistic approach to characterize latent semantic associations among co-occurring words and documents \cite{hofmann1999prob}. By introducing a distribution to reflect the ratings given by each user for each file, pLSA was then applied to predict user preference in collaborating filtering \cite{ekstrand2011collaborative}. To characterize the file requests behavior, pLSA associates a topic among co-occurring files and users, where a topic can be art, children, education, games, or technology, etc. In the following, we introduce pLSA to model the requests of each user.

The joint probability that ${\rm u}_k$ requests ${\rm f}_f$  is
\begin{equation}
\label{joint_p}\textstyle
P({\rm u}_k, {\rm f}_f) = P({\rm u}_k)P({\rm f}_f|{\rm u}_k).
\end{equation}


By introducing latent topic set $\mathcal{Z} = \{{\rm z}_1,{\rm z}_2,...,{\rm z}_{Z} \}$ with $|\mathcal{Z}|=Z$, pLSA associates each possible user request, i.e., ${\rm u}_k \in \mathcal{U}$ requests  ${\rm f}_f \in \mathcal{F}$, with each topic ${\rm z}_j \in \mathcal{Z}$ $(1 \leq j \leq Z)$.  Specifically, the request of each user can be modeled as the following steps with three model parameters:
\begin{itemize}
	\item ${\rm u}_k$ sends a request with probability $P({\rm u}_k)$,
	\item ${\rm u}_k$ chooses a topic ${\rm z}_j$ with probability $P({\rm z}_j|{\rm u}_k)$,
	\item ${\rm u}_k$ prefers  ${\rm f}_f$ in topic ${\rm z}_j$ with probability $P({\rm f}_f|{\rm z}_j)$.
\end{itemize}

The above model is based on the {\em conditional independence} assumption that conditioned on the chosen topic ${\rm z}_j$ by ${\rm u}_k$, ${\rm f}_f$ is chosen independently of ${\rm u}_k$ with probability $P({\rm f}_f|{\rm z}_j)$  rather than $P({\rm f}_f|{\rm z}_j,{\rm u}_k )$, i.e., $P({\rm f}_f |{\rm u}_k ) = \sum_{{\rm z}_j \in {\mathcal{Z}}}P({\rm f}_f |{\rm z}_j) P({\rm z}_j|{\rm u}_k)$.  Then, the joint probability in \eqref{joint_p} can be rewritten as
\begin{equation}
	\label{joint_p_u_f}\textstyle
	\begin{aligned}
	P({\rm u}_k  , {\rm f}_f )& \textstyle = P({\rm u}_k  )P({\rm f}_f |{\rm u}_k )\\
	& \textstyle= P({\rm u}_k ) \sum_{{\rm z}_j \in {\mathcal{Z}}}P({\rm f}_f |{\rm z}_j) P({\rm z}_j|{\rm u}_k).
	\end{aligned}
\end{equation}

According to maximal likelihood principle, we can determine $ P({\rm u}_k ), P({\rm f}_f |{\rm z}_j)$ and $P({\rm z}_j |{\rm u}_k)$ with requests history $n({\rm u}_k , {\rm f}_f)$ by maximizing the log-likelihood function
\begin{equation}
\label{likelihood}\textstyle
\mathcal{L} = \sum_{{\rm u}_k \in \mathcal{U}} \sum_{{\rm f}_f \in \mathcal{F}} n({\rm u}_k , {\rm f}_f) \log P({\rm u}_k , {\rm f}_f ).
\end{equation}


\subsection{Algorithm to Predict User Preferences}

EM algorithm is frequently-used and efficient for maximal likelihood parameter estimation \cite{demp1977maximum}. To exploit EM algorithm, the log-likelihood function in \eqref{likelihood} is rewritten as \cite{hofmann1999prob}
\begin{equation}
\label{likelihood_EM} \textstyle
\mathcal{L} = \sum_{{\rm u}_k \in \mathcal{U}} \sum_{{\rm f}_f \in \mathcal{F}} n({\rm u}_k , {\rm f}_f) \log  \sum_{{\rm z}_j \in {\mathcal{Z}}}  P({\rm z}_j )P({\rm u}_k|{\rm z}_j) P({\rm f}_f |{\rm z}_j),
\end{equation}
which follows by substituting conditional probability $P({\rm z}_j |{\rm u}_k) = {P({\rm u}_k |{\rm z}_j)P({\rm z}_j)}/{P({\rm u}_k)}$ into \eqref{joint_p_u_f}.

Starting from randomly generated initial values for the model parameters  $ P({\rm z}_j )$,  $P({\rm u}_k |{\rm z}_j)$ and $P({\rm f}_f |{\rm z}_j)$, the EM algorithm alternates two steps: expectation (E) step and maximization (M) step. In the E-step, posterior probabilities are computed for latent variable ${\rm z}_j$ with current estimation of the parameters as
\begin{equation}
\label{E_step}\textstyle
P({{\rm z}_j|{\rm u}_k,{\rm f}_f})= \frac{P({\rm z}_j )P({\rm u}_k|{\rm z}_j) P({\rm f}_f |{\rm z}_j)}{\sum_{{{\rm z}_{j'}} \in \mathcal{Z}} P({\rm z}_{j'} )P({\rm u}_k|{\rm z}_{j'}) P({\rm f}_f |{\rm z}_{j'})},
\end{equation}
which is the probability that ${\rm f}_f$ requested by  ${\rm u}_k$ belongs to topic ${\rm z}_j$. In the M-step, given $P({{\rm z}_j|{\rm u}_k,{\rm f}_f})$ computed by previous E-step, the parameters are updated  as \cite{hofmann1999prob}
\begin{subequations}
	\begin{align}
	& \textstyle P({\rm z}_j) =  \frac{ \sum_{{\rm f}_f \in \mathcal{F}}  \sum_{{\rm u}_k \in \mathcal{U}}  n({\rm u}_k,{\rm f}_f) P({\rm z}_j|{\rm u}_k,{\rm f}_f)    }
	{  \sum_{{\rm f}_f \in \mathcal{F}} \sum_{{\rm u}_k \in \mathcal{U}}     n({\rm u}_k,{\rm f}_f)   }, \label{M-1}\\
	& \textstyle P({\rm u}_k|{\rm z}_j) = \frac{ \sum_{{\rm f}_f \in \mathcal{F}}  n({\rm u}_k,{\rm f}_f)   P({\rm z}_j|{\rm u}_k,{\rm f}_f)    }
	{  \sum_{{\rm f}_f \in \mathcal{F}} \sum_{{\rm u}_{k'} \in \mathcal{U}}    n({\rm u}_{k'},{\rm f}_f)   P({\rm z}_j|{\rm u}_{k'},{\rm f}_f)   },\label{M-2} \\
	& \textstyle P({\rm f}_f|{\rm z}_j) = \frac{ \sum_{{\rm u}_k \in \mathcal{U}}  n({\rm u}_k,{\rm f}_f)  P({\rm z}_j|{\rm u}_k,{\rm f}_f)    }
	{   \sum_{{\rm u}_k \in \mathcal{U}} \sum_{{\rm f}_{f'} \in \mathcal{F}}   n({\rm u}_k,{\rm f}_{f'})  P({\rm z}_j|{\rm u}_k,{\rm f}_{f'}) }.\label{M-3}
	\end{align}
\end{subequations}

Using the EM algorithm, the joint distribution $P({\rm u}_k , {\rm f}_f )= \sum_{{\rm z}_j \in {\mathcal{Z}}}  P({\rm z}_j )P({\rm u}_k|{\rm z}_j) P({\rm f}_f |{\rm z}_j)$ can be estimated. Then, we can predict the request probability of ${\rm u}_k$ and the preference of ${\rm u}_k$ for ${\rm f}_f$ as $\hat{w}_k = P({\rm u}_k) = \sum_{{\rm f}_f \in \mathcal{F}}P({\rm u}_k , {\rm f}_f )$ and $\hat{q}_{k,f}= P({\rm f}_f|{\rm u}_k) = P({\rm u}_k , {\rm f}_f )/P({\rm u}_k) $, respectively.
The detailed steps for predicting $\bf {w}$ and $\bf {Q}$ are provided in Algorithm \ref{Algor_2}.

\begin{algorithm}[!htb]
	\caption{Learning user preferences based on pLSA.}
	\label{Algor_2}
	\begin{algorithmic}[1] 
		\REQUIRE ~
		 $\bf N $; $Z$;
		Stop condition $0 <\epsilon < 1$;  \\
		Initialize:
		$ P^{(0)}({\rm z}_j)$ ,  $P^{(0)}({\rm u}_k |{\rm z}_j) $ and $P^{(0)}({\rm f}_f |{\rm z}_j)$; Step $i \leftarrow 1$; Difference $\Delta \leftarrow \infty$;
		Log likelihood $\mathcal{L}(0)\leftarrow0$;
		
		\WHILE{$\Delta > \epsilon$}
		\STATE Using $ P^{(i-1)}({\rm z}_j)$,  $P^{(i-1)}({\rm u}_k |{\rm z}_j)$ and $P^{(i-1)}({\rm f}_f |{\rm z}_j)$  in \eqref{E_step} to compute $P^{(i)}({\rm z}_j|{\rm u}_k,{\rm f}_f)$;
		\STATE Using $P^{(i)}({\rm z}_j|{\rm u}_k,{\rm f}_f)$ in \eqref{M-1} \eqref{M-2} and \eqref{M-3} to compute $ P^{(i)}({\rm z}_j)$, $P^{(i)}({\rm u}_k |{\rm z}_j) $ and $P^{(i)}({\rm f}_f |{\rm z}_j)$;
		\STATE Compute log likelihood $\mathcal{L}(i)$ with $ P^{(i)}({\rm z}_j)$,  $P^{(i)}({\rm u}_k |{\rm z}_j) $ and $P^{(i)}({\rm f}_f |{\rm z}_j)$ by \eqref{likelihood_EM};
		\STATE $\Delta = |\mathcal{L}(i) - \mathcal{L}(i-1)  | $;	
		\STATE $i \leftarrow i + 1$;
		\ENDWHILE
		\STATE $P({\rm u}_k , {\rm f}_f ) \leftarrow \sum_{{\rm z}_j \in {\mathcal{Z}}}  P^{(i)}({\rm z}_j )P^{(i)}({\rm u}_k|{\rm z}_j) P^{(i)}({\rm f}_f |{\rm z}_j)$;
		\STATE $\hat{w}_k  \leftarrow \sum_{{\rm f}_f \in \mathcal{F}}P({\rm u}_k , {\rm f}_f )$ to compute $\bf \hat{w}$;
		\STATE $\hat{q}_{k,f} \leftarrow  P({\rm u}_k , {\rm f}_f )/P({\rm u}_k) $ to compute $\bf \hat{Q}$;
		\ENSURE $\bf \hat{w}$ and $\bf \hat{Q}$.
	\end{algorithmic}
\end{algorithm}


\section{Simulation Results}
\label{sec:simulation}
In this section,  we demonstrate the offloading gain introduced by the caching policy exploiting {\em perfect} and {\em predicted} user preferences over that with {\em perfect} and {\em predicted} content popularity. Specifically, we compare the following schemes:
\begin{enumerate}
\item ``\textbf{S1-perfect}'': The proposed caching policy with perfect user preference, which is the solution of problem \textbf{P1}.
\item ``\textbf{S2-perfect}'': The existing caching policy optimized with perfect content
popularity, which is the solution of problem \textbf{P2}.
\item ``\textbf{S1-pLSA}'': The proposed caching policy with predicted user preference, where Algorithm 2 is used to predict $\bf \hat{w}$ and $\bf \hat{Q}$, and then Algorithm 1 is used to find the caching policy.
\item ``\textbf{S2-pLSA}'': The existing caching policy with predicted content popularity, which is obtained with Algorithm 2 as $\hat{p}_f  = \sum_{{\rm u}_k \in \mathcal{U}}P({\rm u}_k , {\rm f}_f )$ via \eqref{p_f}.
\item ``\textbf{S1-baseline}'': The proposed caching policy with predicted user preference, which is obtained by extending the frequency-count popularity prediction method in  \cite{tatar2014survey} as $\hat{q}_{k,f} = \frac{n({\rm u}_k, {\rm f}_f)}{\sum_{f=1}^{F} n({\rm u}_k, {\rm f}_f)}, \hat{w}_k =  \frac{ \sum_{f=1}^{F} n({\rm u}_k, {\rm f}_f) }{\sum_{k'=1}^{K} \sum_{f=1}^{F} n({\rm u}_{k'}, {\rm f}_f)}$.
\item ``\textbf{S2-baseline}'': The existing caching policy with predicted content popularity, which is obtained by the learning algorithm in \cite{Blasco14learning}.
\end{enumerate}

We employ cross validation to demonstrate the learning performance of pLSA, i.e.,  we do not use pLSA model presented in Section \ref{sec:predict} but use the synthetic model introduced in Section II  to generate user preferences in simulation.

We consider a square cell with side length $500$ m, where $K=100$ users are  uniformly located.
The file catalog size $F=500$, and each user is willing to cache $M=5$ files holding $1\%$ of all files. The parameter of Zipf distribution is $\beta =0.6$, which is slightly smaller for small region such as campus than that observed at the Web proxy  as reported in \cite{gill2007youtube}. We divide time into two-hour periods.\footnote{Using other values as the duration of each period does not affect the predicting performance of user preferences, since the performance only depends on the number of requests in the past. Yet the period can not be too short, since a frequent caching placement brings extra traffic load.} The request arrival rate of the users in the cell is $0.4$ requests per second, which reflects the high traffic load scenario in \cite{3GPP.PT}. The cached files at each user are updated in off-peak time.

%

\begin{figure}[!htb]
	\centering
	\includegraphics[width=0.5\textwidth]{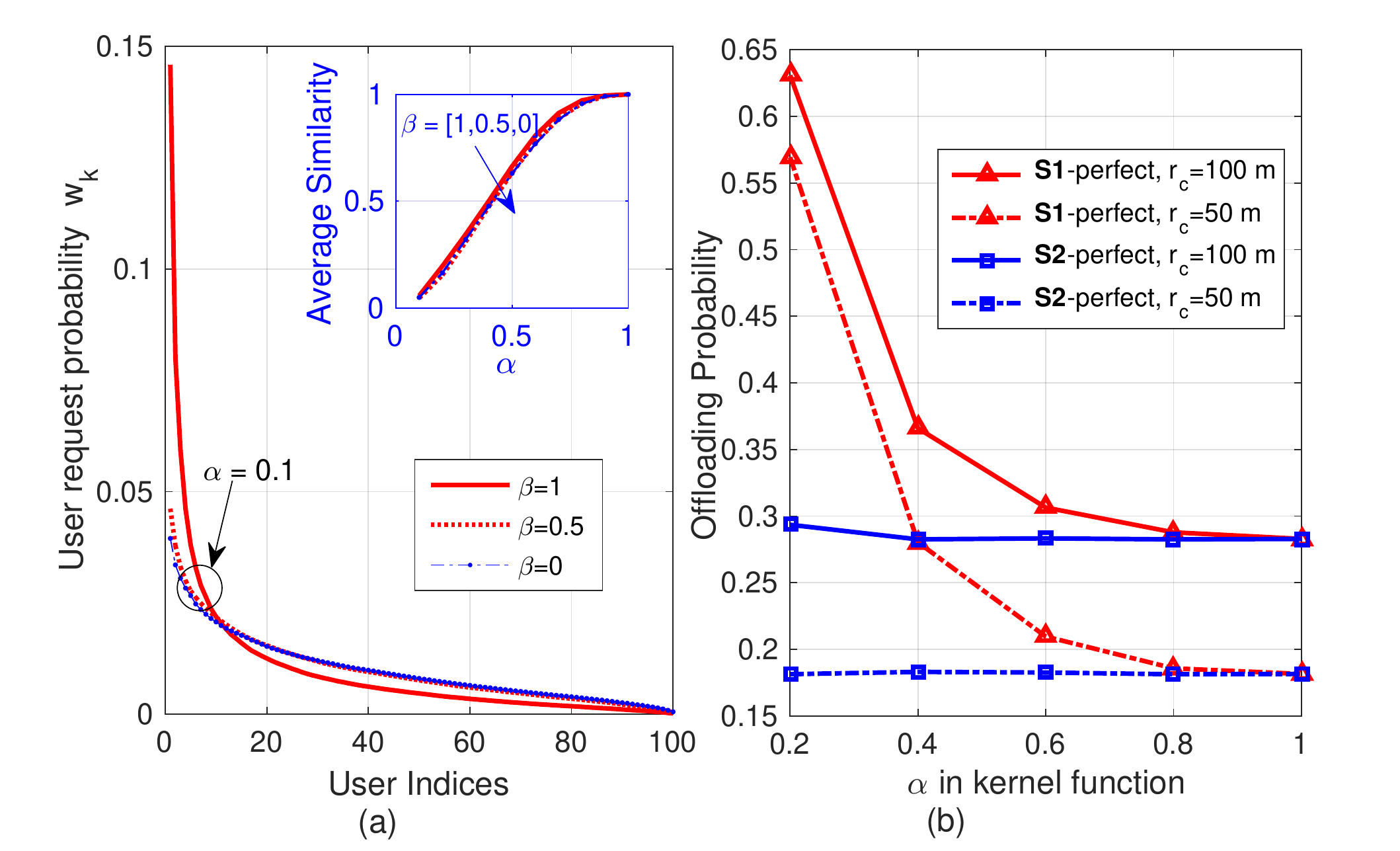}\\
	\caption{User request probability, average cosine similarity and offloading probability with perfectly known user preferences and content popularity. }\label{fig.1}
\end{figure}

In Fig. \ref{fig.1}(a), we first show the impact of $\alpha$ in the kernel function. We can see that the synthetic
user preference model can capture different levels of similarity among user preferences by adjusting $\alpha$, while the Zip parameter $\beta$ has negligible impact on the average cosine similarity. This seems counter-intuitive, since a more skewed popularity distribution is believed to imply highly similar user preferences. However, such an intuition comes from the implicit assumption that the users sent their requests with equal probability, which is not true in reality, as reported from recent big data analysis. From the figure we can observe that $\beta=1$ even when $\alpha =0.1$, because a few users in the cell send file requests with high probability, i.e., only a few users are very active.

In Fig. \ref{fig.1}(b), we show the impact of $\alpha$ and $r_{\rm c}$ on offloading probability achieved by ``{\textbf S1}-perfect'' and ``{\textbf S2}-perfect''. We can see that offloading gain can be remarkably improved by using the proposed caching policy when $\alpha$ is small. This suggests that optimizing caching policy according to user preferences is critical when the user  preferences are less correlated. As expected, when $\alpha \to 1$, the performance of the two policies almost coincide. We can also see that the offloading gain can be improved by extending collaboration distance, but the gain by using the proposed policy reduces as indicated in Remark 3. This is because, with the growth of $r_{\rm c}$, the number of users to which a helper user can share cached files increases, and thus caching policy needs to consider more user preferences.

\begin{figure}[!htb]
	\centering
	\includegraphics[width=0.5\textwidth]{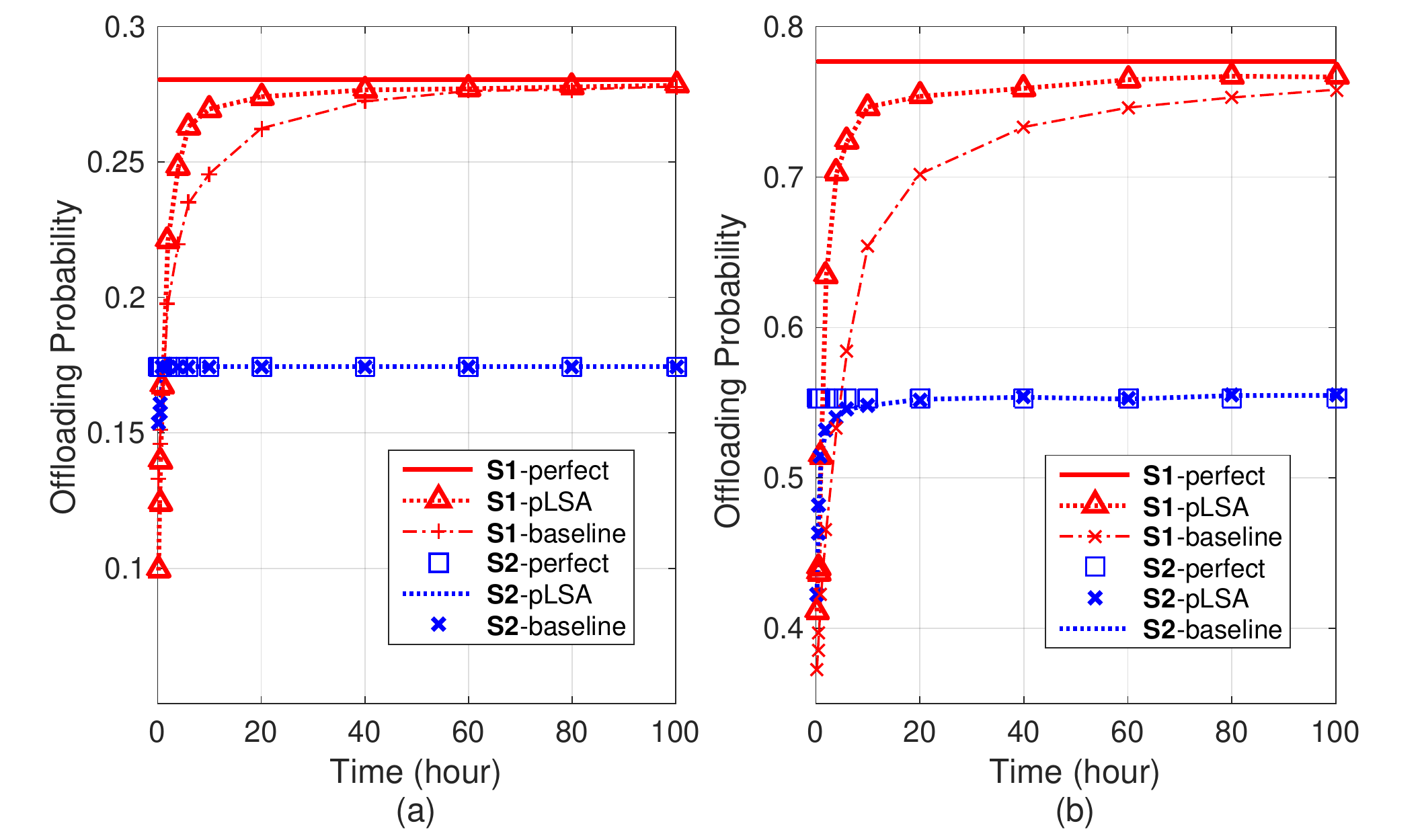}\\
	\caption{Time evolution of offloading probability with different algorithms, $\alpha = 0.4$, $r_{\rm c} = 50$ m, $Z=10$ for pLSA, in (a) $M = 5$ and in (b) $ M = 50$. }\label{fig.2}
\end{figure}

In Fig. \ref{fig.2}, we show the offloading probability achieved by the six schemes during the learning procedure. The curves for \textbf{S2} almost overlap. Compared to the proposed caching policy with predicted user preference, we can see that \textbf{S2} with predicted content popularity can approach the performance of  \textbf{S2} with perfect prediction more quickly. This is because, with given requests history, the number of requests for each file from each user is much less than that from all users. As a result of the data sparsity,  predicting user preference is more difficult than predicting content popularity. Nonetheless, the proposed caching policy with predicted user preference can quickly achieve high offloading gain over \textbf{S2} with predicted (and even perfect) content popularity. We can also see that the proposed caching policy with pLSA is superior to that with the baseline, especially in the initial stage of the learning procedure and/or when the storage size at each user $M$ is large. This is because some unpopular files can be cached with large $M$. For the unpopular files, the number of accumulated requests is less and the preference prediction is more difficult.

\section{Conclusions}
In this paper, we optimized the caching policy by learning user preference for cache-enabled D2D communications. We first formulated an optimization problem with given user preferences to maximize the offloading probability, which was shown as NP-hard. A greedy algorithm was proposed to solve the problem. Then, we modeled the user request behavior by pLSA, based on which the EM algorithm was used to predict the user preference and file request probability. Simulation results showed that using pLSA and EM algorithm can quickly learn individual user behavior. Compared to existing popularity-driven caching policy, the performance can be remarkably improved by the proposed caching policy, especially when the user preferences are less correlated.

\label{sec:conclusion}
\bibliographystyle{IEEEtran}
\bibliography{CBQ_J}
\end{document}